\documentclass[referee]{aa} 
\usepackage{graphicx}
\usepackage{txfonts}
\usepackage{longtable}    
\usepackage{aalongtable}    
\usepackage[authoryear]{natbib}
\bibpunct{(}{)}{;}{a}{}{,} 
\newcommand{\kms}{km\,s$^{-1}$}

\begin{document}
%
   \title{Lithium abundance in  the metal-poor open cluster NGC~2243 
\thanks{Based on observations collected at the ESO VLT, Paranal Observatory (Chile). }}

\author{P. Fran\c{c}ois \inst{1,2}  \and L. Pasquini \inst{3} 
\and K. Biazzo \inst{4,3}
\and P. Bonifacio \inst{1} \and R. Palsa \inst{3}}
\offprints{P. Fran\c{c}ois}
\mail{patrick.francois@obspm.fr}

\institute{
 GEPI - Observatoire de Paris, 64 Avenue de l'Observatoire, 75014 Paris, France
\and UPJV - Universit\'e de Picardie Jules Verne, 80000 Amiens, France
\and ESO - European Southern Observatory, Karl-Schwarzschild-Str. 3, 85748 Garching bei M\"unchen, Germany
\and INAF - Capodimonte Astronomical Observatory, via Moiariello 16, 80131 Naples, Italy
}

   \date{Received December 19, 2012; Accepted ...}
\abstract
{Lithium is a fundamental element for studying the mixing mechanisms acting in the stellar interiors, for understanding the chemical evolution of 
the Galaxy and the Big Bang nucleosynthesis. The study of Li in stars of open clusters (hereafter OC) allows a detailed comparison with stellar evolutionary models and permits us to trace its galactic evolution. The OC NGC~2243 is particularly interesting because of its low metallicity ([Fe/H]=$-0.54 \pm0.10$ dex).}
{We measure the iron and lithium abundance in stars of the metal-poor OC NGC~2243.  The first aim is to determine 
whether the Li dip extends to such low metallicities, the second is to compare the results of our Li analysis in this OC with those present in 
47 Tuc, a globular cluster of similar metallicity.} 
{We performed a detailed analysis of high-resolution spectra obtained with the multi-object facility FLAMES at the ESO 
VLT 8.2m telescope. Lithium abundance was derived through line equivalent widths and the OSMARCS atmosphere models. 
 Iron abundances from  $\ion{Fe}{i}$ and $\ion{Fe}{ii}$ lines  have also been  measured and used to check  the atmospheric model parameters.  } 
{The Li line is detected in 27 stars confirmed as likely cluster members by repeated radial velocity measurements. We determine a
Li dip center of 1.06 $M_\odot$, which is much smaller than that observed in solar metallicity and metal-rich clusters. This finding confirms and strengthens 
the conclusion that the mass of the stars in the Li dip strongly depends  on stellar metallicity. The mean Li abundance of the cluster is  $\log n{\rm (Li)}=2.70$ dex, which is substantially higher than that observed in 47 Tuc. We estimated 
an iron abundance of [Fe/H]=$-0.54 \pm0.10$ dex for NGC~2243, which is similar (within the errors) to previous findings.  
 {{The [$ \alpha$/Fe] content ranges from $0.00\pm0.14$ for Ca to $0.20\pm0.22$ for Ti,}}
which is low when compared to thick disk stars and to Pop~II stars, but compatible with thin disk objects. We found a mean radial velocity of 61.9 $\pm$ 0.8 \kms   for the cluster. }
{We confirm a correlation between the Li dip cool-side position in mass as a function of the cluster's [Fe/H].
The Li abundance found in the metal-poor OC NGC2243 agrees well with results obtained  for 
the interstellar medium in the Small Magellanic Cloud (SMC) having similar low metallicity. This  value of Li is comparable to the primordial Li abundance deduced from WMAP measurements, therefore putting strong constraints on the models of Li enrichment during the early history of our Galaxy. }
   
\titlerunning{Lithium in NGC~2243}
\authorrunning{Fran\c{c}ois et al. }
\maketitle

%

%
%

\section{Introduction}
The measurement of the Li abundance of turnoff and dwarf stars in OCs of different ages and metallicities is a very powerful 
tool that can be used to investigate a number of topics: the possibility of sampling stars of different evolutionary status and original mass allows to
understand how this element is destroyed and to test different evolutionary models and mixing mechanisms 
(Boesgaard \& Tripicco 1986; Pasquini 2000; Randich et al. 2000; Randich et al. 2003; Smiljanic et al. 2010). 
This is possible because for OC stars we can obtain reliable ages, metallicities, masses, and evolutionary stages from precise 
photometry and high-resolution spectroscopy. Cluster stars are the most powerful targets that can be studied to disentangle the effect of each of these parameters 
on mixing phenomena. Since Li is burnt through $(p,\alpha)$ reactions at a temperature of $\simeq 2.5 \times 10^{6}$ K, differences in 
Li abundance in stars of different masses and temperatures can be used to constrain the physical processes that bring the surface 
material to the interior of the star. 

The cluster we have studied, NGC~2243, is particularly interesting because it is one of the most metal-poor OCs accessible for 
high-resolution spectroscopy. It is almost as metal-poor as the globular cluster 47 Tucanae, considered  ``metal rich'' among  globular clusters with a metallicity of [Fe/H]=-0.76 $\pm$ 0.04 dex (\citet{Koch08}). 
The comparison of the Li abundances in the two clusters is very interesting, since, in spite of the similar 
metallicity, NGC~2243 is much younger than 47 Tuc. The so-called ``Li dip'', i.e., the disappearance of Li in the spectra of stars over 
a narrow temperature range, is a feature that is present in OCs, but in none of the globular clusters. It is interesting  
to investigate whether it is present also in an OC with a low metallicity comparable to that of ``metal-rich'' globular clusters. 
Another interesting characteristic of OCs  that are young enough to have hot main sequence stars on the blue side of the dip (see, e.g., \citealt{Sestito05}, and references therein) is that hot stars out of the ``Li dip'' show a high level of 
Li abundance, close to the abundance present in the interstellar medium when they formed. This  
initial Li abundance in metal-poor clusters will provide precious data to be compared with 
models of Li evolution in the Galaxy (e.g., Matteucci 2010)  and primordial nucleosynthesis \citep{CFO08}.

The first determination of the Li abundance in this cluster was performed by Hill \& Pasquini (2000). They found a mean 
Li abundance of $\log n{\rm (Li)}=2.35$ in the four stars for which they could detect Li. These authors hinted at the presence of a Li 
dip, but could not confirm it, given their limited sample. 

Here, we present a new study of elemental abundances in NGC~2243, based on high-resolution spectra obtained with FLAMES on the VLT. 
In Sect.~\ref{sec:ngc2243} we briefly describe the cluster. The observations and data reduction are given in Sect.~\ref{sec:Obs}, 
while the data analysis is discussed in Sect.~\ref{sec:analysis}. The results, discussion, and conclusions are presented in 
Sects.~\ref{sec:results}, \ref{sec:discussion}, and \ref{sec:concl}.

\label{sec:Intro}

%
%

\section {NGC 2243: an old metal-poor open cluster}
\label{sec:ngc2243}

NGC~2243 is an old and metal-poor OC located towards the anti-center at $R_{\rm G}=10.76$ kpc and $z=1.1$ kpc away from the plane 
of the Milky Way ($l_{II}=239\degr.478$, $b_{II}=-18\degr.014$). The first CCD photometry of NGC~2243 was obtained by Bonifazi et al. (1990) 
who found an age of $4\pm1$ Gyr, a metallicity of [Fe/H]$=-0.80\pm0.10$, and a distance modulus $(m-M)_{0}=12.8\pm0.2$. From $BV$ CCD 
photometry, Bergbusch et al. (1991) derived a distance modulus of 13.05 (assuming $E(B-V)=0.06$), a metallicity of $-0.47$ dex, an 
overabundance of oxygen [O/Fe]=+0.23, and an age of 5 Gyr. 
Gratton \& Contarini (1994)  confirmed this low metallicity using high-resolution spectra. 
They determined  the iron abundance in two giants of the cluster and found [Fe/H]$=-0.48\pm0.15 $ dex 
and  [Ca/Fe]$=0.18 \pm0.17 $ dex.
Fitting the data of Bergbusch et al. (1991) to new isochrones, assuming core overshooting, 
Vandenberg et al. (2006) obtained [Fe/H]$=-0.61$, [$\alpha$/Fe]=0.3, an age of 3.1 Gyr and $(m-M)_{V}=13.15$, assuming $E(B-V)=0.062$. 
From $VI$ CCD photometry, Kaluzny et al. (1996) found $E(V-I)=0.10\pm0.04$. Kaluzny et al. (2006) extended their study of NGC~2243 to 
the analysis of detached eclipsing binaries. They derived a distance modulus $(m-M)_{V}=13.24\pm0.08$. Using model age-luminosity and 
age-radius relations for the binary system NV CMa, a member of the cluster, they obtained an age of $4.35\pm0.25$ Gyr assuming a metallicity 
[Fe/H]$=-0.525$. However, these values are sensitive to the adopted metallicity, a change of $-0.08$ dex in the metallicity leading to 
an age decrease of $\simeq 0.5$ Gyr. 

%
%
\section{Observations and data reduction}
\label{sec:Obs}
The spectra were obtained in four hours in four observing nights during the FLAMES+GIRAFFE Science Verification program. 
FLAMES (Pasquini et al. 2002) is mounted at the Nasmyth A platform of 
the 8.2m UT2/Kueyen of the VLT. Observations were carried out in MEDUSA mode with two GIRAFFE high-resolution (HR) settings  and with a single 
2K$\times$4K EEV CCD (15 $\mu$m pixels). In particular, the employed settings were Filter N. 14 ({\it \'echelle} grating order 
N. 9) with central wavelength 651.5 nm and wavelength range between 638.3 and 662.6 nm, and Filter N. 15 ({\it \'echelle} order 
N. 8) with the central wavelength 679.7 nm and wavelength range between 659.9 and 695.5 nm. 
The resolution was $R\sim$28\,800 and $R\sim$19\,300, respectively. The log of the observations is given in Table~\ref{tab:observations}. 

We selected from the catalog of \cite{Kalu1996} the main-sequence (MS), turnoff (TO), and (sub-)giant stars with $14\fm0\le V \le17\fm5$ and, 
thanks to the GIRAFFE configuration, we were able to observe 100 targets (Tables~\ref{tab:LiAb}, \ref{tab:LiAbUp}, \ref{tab:binaries}). 
Separate exposures were obtained to be able to identify short and intermediate period binaries by comparing the radial velocities 
at different epochs. The spectra have a typical signal-to-noise ($S/N$) ratio of 30--40. 

The observations were reduced using the version 1.12 of the GIRAFFE Base-Line Data Reduction Software 
(girBLDRS\footnote{http://girbldrs.sourceforge.net/}; \citealt{Blecha2000}), which is a set of python 
scripts/modules and a C library. With it we removed the instrumental signature from the observed data, 
subtracting the bias and dividing by the normalized flat-field. Flat-field acquisitions were also 
used to trace the position of all the fibers, and to derive the parameters for the optimal extraction 
of the science exposures. Finally, the wavelength calibration was obtained using the day-time Th-Ar lamp exposure. 

Radial velocities were measured using the IRAF\footnote{IRAF is distributed by the National Optical Astronomy 
Observatory, which is operated by the Association of the Universities for Research in Astronomy, Inc. (AURA) under 
cooperative agreement with the National Science Foundation.} package FXCOR, which cross-correlates the observed 
spectrum with a template. As a template, we used a solar spectrum acquired with FLAMES/GIRAFFE. Finally, the 
heliocentric correction was applied. We considered only the spectra acquired with the HR14 setting, as it leads to 
the radial velocity measurements with the lowest rms uncertainties (around 1 \kms),  compared to the 2.5 \kms obtained with the HR15 setting. The two spectra/star/setting were finally co-added to perform the spectroscopic determination of elemental abundance (and effective temperature). 

The GIRAFFE solar spectra\footnote{http://www.eso.org/observing/dfo/quality/GIRAFFE/pipeline/solar.html}, 
taken with the same setups of our observations, are used throughout this work for spectroscopic comparison with the stars 
and the synthetic spectra. Each solar spectrum was obtained by averaging most of the GIRAFFE 
spectra (some spectra show clear flat field problems and were not used) and it has a nominal $S/N$ ratio above 400.
 
\begin{table}  
\caption{Log of the observations.}
\label{tab:observations}
\begin{center}  
\begin{tabular}{cccccc}
\hline
\hline
$\alpha$    & $\delta$  &  Date      &  UT      & $t_{\rm exp}$ & Filter \\
(\degr)     &  (\degr)  & (d/m/y)    & (h:m:s)  &  (s)          &        \\ 
\hline
97.40327     & $-$31.28795   & 27/01/2003 & 04:12:26 & 3600	& HR15	 \\
97.40331     & $-$31.28800   & 28/01/2003 & 05:49:49 & 3600	& HR14	 \\
97.40326     & $-$31.28808   & 29/01/2003 & 03:29:14 & 3600	& HR15     \\
97.40333     & $-$31.28806   & 31/01/2003 & 02:14:15 & 3600	& HR14     \\
\hline
\end{tabular}
\end{center}
\end{table}  


%
%

\section{Data analysis}
\label{sec:analysis}

\subsection{Radial velocities}
\begin{figure}
\centering
\includegraphics[width=0.5\textwidth]{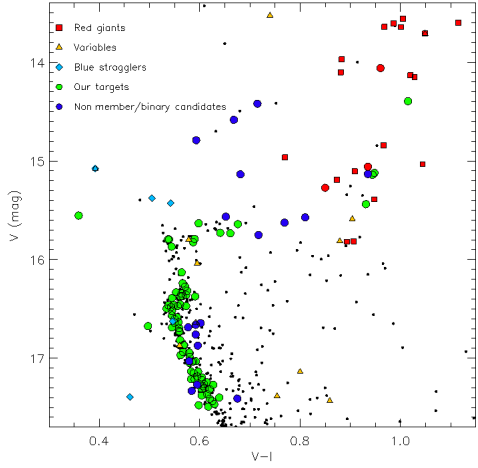}
\caption{Portion of the CMD of NGC~2243. Small dots represent $V$ and $V-I$ data taken from \cite{Kalu1996}, 
while squares, triangles, and diamonds mark the positions of red giants, variables, and blue stragglers 
taken from the literature (WEBDA database). Circles refer to our targets, where 
the position of the most-probable single stars and binaries or non-members 
(from our radial velocity analysis) are marked in green and blue, 
respectively, while red giants from the literature (WEBDA database) are in red.} 
\label{fig:cmd_ngc2243}
\end{figure}

The evolutionary status of the analyzed stars is indicated by their position in the color-magnitude diagram (CMD) shown in 
Fig.\ref{fig:cmd_ngc2243}, in which our sample is shown with circles. The stars are mostly MS, TO, post-TO, subgiants, 
and giants, where the region around the TO is very well sampled. 

\begin{figure}
\centering
\includegraphics[width=0.5\textwidth]{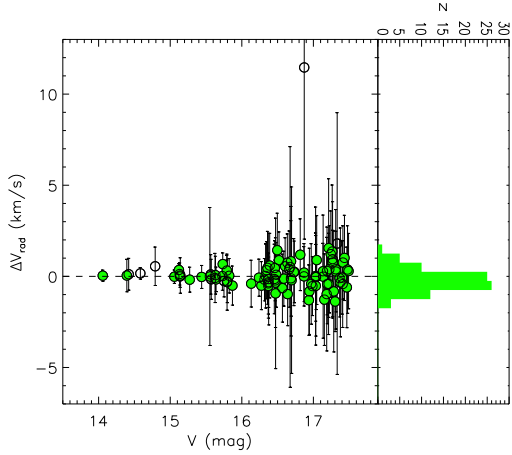}
\caption{Final radial velocity variations versus $V$ magnitude. The right part of the figure represents the histogram of the computed mean RV differences. 
Open symbols refer to the final selection of binaries or non-members.} 
\label{fig:delta_vrad_V}
\end{figure}

\begin{figure}
\centering
\includegraphics[width=0.5\textwidth]{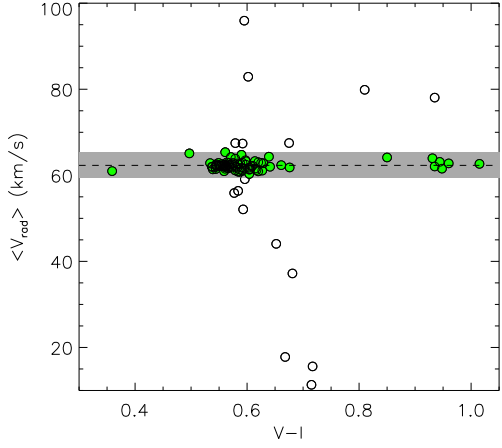}
\caption{Mean radial velocities as a function of the $V-I$ color. The dashed line represents the final mean RV $\sim 62$ \kms, 
while the filled area is the 
$\pm 3 \sigma$ level ($\sim 3$ \kms). Open symbols refer to the final selection of binaries or non-members.} 
\label{fig:vrad_VI}
\end{figure}

\begin{figure}
\centering
\includegraphics[width=0.5\textwidth]{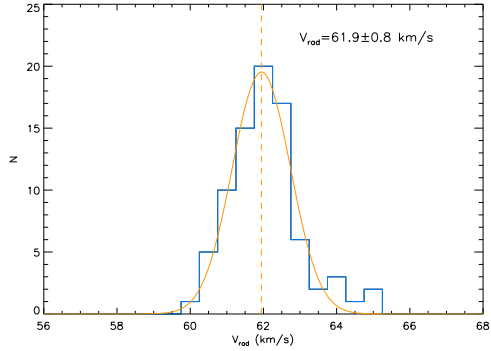}
\caption{Histogram of the radial velocity distribution of the 82 most-probable single members selected in NGC~2243 (continuous line). 
A Gaussian fit to the distribution of member stars  is also displayed, with a mean 
value of $<V_{\rm rad}>=61.9\pm0.8$ \kms (dashed line).} \label{fig:vrad_distribution_fin_selNEW}
\end{figure}

Having several exposures for each object allows the comparison of the single radial velocity (RV) for each star. 
We have retained as most-probable single members all those showing RV variations smaller than 2 \kms~  
in the two exposures acquired and having a mean velocity within about 3 sigma ($\approx 3$ \kms) from the median 
cluster RV. The final values of the difference in the computed RV is given in Fig.~\ref{fig:delta_vrad_V}, 
as well as the distribution, which is well represented by a Gaussian with a $\sigma$ of 0.8 \kms. The final mean
 radial velocities are shown in Fig.~\ref{fig:vrad_VI} as a function of the $V-I$ color. Most of the RV 
are concentrated at around 62 \kms, close to the expected radial velocity of the cluster. A total of 82 stars are retained as {\it bona fide} RV single members because their RV distribution can be well approximated by a Gaussian with central RV of $61.9\pm0.8$ \kms  as shown in Fig.~\ref{fig:vrad_distribution_fin_selNEW}. In Tables~\ref{tab:LiAb}, \ref{tab:LiAbUp}, and \ref{tab:binaries} the RV values of 
the stars with well-measured Li abundances, upper-limited Li abundances, and most-probable binaries
 or non-members are listed.

%
%

%
%

\subsection{Atmospheric parameters}
We used the photometric results for $V-I$ colors from \cite{Kalu1996} to derive the effective temperatures
 from the relations of \cite{Casa2010}. We also attempted to measure effective temperatures from 
 our GIRAFFE spectra, using the H$\alpha$ wings, but the results were not convincing when comparing 
these effective temperatures with photometry. Most likely the flat field procedure adopted did not
 succeed in fully removing the rather strong blaze present in the spectra, possibly because of the relatively
  close position of the H$\alpha$ line to the spectrum edge. 

The surface gravities for the program stars were obtained from the photometric data quoted above and computed using the standard relation given in 
Eq.~\ref{eq:logg}. We adopted the following solar values: $\log g_{\odot} = 4.44 dex, T_{\rm eff, \odot} = 5790$ K, $M_{\rm bol, \odot} = 4.72$ mag. 
The bolometric corrections were computed using the relation of \cite{Alon1999}. We assumed a TO mass of $1.15 M_{\odot}$ and a distance modulus of 13.15. We note that even if the adopted turnoff mass  was wrong by 25$\%$, the errors in computing $\log g$ would be less than 0.1 dex.

\begin{equation}
\label{eq:logg}
 \log g_{*} = \log g_{\odot} + \log \frac{M_{*} }{M_{\odot}} + \log \frac {T_{\mathrm{eff},*}}{T_{\mathrm{eff},\odot}} + 0.4 (M_{\mathrm{bol},*} - M_{\mathrm{bol},\odot}). 
\end{equation}
The microturbulence velocities were computed using the relation of \cite{Edva1993}.

%
%

\section{Results}
\label{sec:results}

\subsection{Lithium abundance}
The equivalent widths of the Li resonance doublet at $\lambda=$670.78 nm was measured using the code 
{\it FITLINE} written by P. Fran\c cois. Details on the algorithm can be found in \cite{Lema2007}. The line was 
measurable in 27 stars. When the lines were not detectable, an upper limit for the equivalent width of the line was 
established via the Cayrel formula \citep{Cayrel1988} and translated into an upper limit for the Li abundance. 

We then derived Li abundances for the measured line equivalent widths  using {\it Turbospectrum} 
\citep{Alva1998} taking the most updated version of the OSMARCS atmosphere models  
(e.g., \citealt{Gusta2008}, \citealt{Plez2008}). Plane parallel 1D models have been used for the computations. We adopted the 
Grevesse \& Sauval (1998) solar abundances. 
In Table~\ref{tab:LiAb} the list of stars for which the Li abundances could be measured is given, 
while Table~\ref{tab:LiAbUp} lists the estimation of the 
Li abundance upper limits for the remaining stars of the  sample.

\begin{table*}
\scriptsize
\caption{Stars with well-measured lithium abundance.}
\label{tab:LiAb}
\begin{center}  
\begin{tabular}{lcccccccccc}
\hline
Object & $V$ & $V-I$ & $T_{\rm eff}$ & $\log g$ & [\ion{Fe}{i}/H] & [\ion{Fe}{ii}/H] &$\log n{\rm (Li)}$ & $V_{\rm rad}^{1}$ & $V_{\rm rad}^{2}$ & $<V_{\rm rad}>$\\
       & (mag) &  & (K) &  &  &  & & (\kms)& (\kms)& (\kms)\\
\hline

 148 &	17.037	& 0.504 & 6646 & 4.39 &	$-$0.56 &	$-$0.62 &	2.46  & 60.8$\pm$1.4& 60.9$\pm$1.2& 60.8$\pm$0.9\\
 231 &	17.400	& 0.559 & 6355 & 4.45 &	$-$0.64 &	$-$0.62 &	2.44  & 64.3$\pm$1.3& 64.4$\pm$1.5& 64.3$\pm$1.0\\
 244 &	17.467	& 0.549 & 6405 & 4.49 &	$-$0.61 &	        &	2.64  & 62.5$\pm$1.7& 63.1$\pm$1.5& 62.8$\pm$1.1\\
 403 &	14.394	& 0.935 & 5051 & 2.77 &	$-$0.33 &	$-$0.42 &	1.51  & 62.7$\pm$0.8& 62.6$\pm$0.5& 62.7$\pm$0.4\\
 414 &	17.412	& 0.540 & 6452 & 4.48 &	$-$0.59 &	$-$0.54 &	2.71  & 63.4$\pm$1.6& 62.7$\pm$1.2& 63.0$\pm$1.0\\
 655 &	16.969	& 0.482 & 6773 & 4.39 &	$-$0.56 &	$-$0.59 &	2.66  & 61.9$\pm$1.9& 62.3$\pm$1.1& 62.1$\pm$1.1\\
 728 &	16.987	& 0.500 & 6668 & 4.37 &	$-$0.55 &	$-$0.40 &	2.58  & 61.8$\pm$1.0& 62.3$\pm$1.3& 62.0$\pm$0.8\\
 766 &	17.315	& 0.540 & 6452 & 4.44 &	$-$0.55 &	        &	2.47  & 61.1$\pm$1.7& 60.8$\pm$1.4& 60.9$\pm$1.1\\
 873 &	17.377	& 0.524 & 6536 & 4.49 &	$-$0.58 &	$-$0.63 &	2.55  & 60.2$\pm$2.5& 60.5$\pm$1.6& 60.3$\pm$1.5\\
 1106&	17.477	& 0.518 & 6568 & 4.54 &	$-$0.50 &	        &	2.73  & 63.5$\pm$1.9& 63.1$\pm$1.5& 63.3$\pm$1.2\\
 1161&	17.194	& 0.524 & 6536 & 4.42 &	$-$0.76 &	$-$0.87 &	2.33  & 61.6$\pm$1.6& 62.0$\pm$1.5& 61.8$\pm$1.1\\
 1183&	17.213	& 0.510 & 6612 & 4.45 &	$-$0.43 &	        &	2.83  & 62.5$\pm$3.0& 61.0$\pm$2.3& 61.8$\pm$1.9\\
 1189&	17.366	& 0.535 & 6478 & 4.47 &	$-$0.56 &	        &	2.57  & 61.5$\pm$1.5& 61.6$\pm$1.5& 61.6$\pm$1.1\\
 1241&	15.824	& 0.507 & 6629 & 3.90 &	$-$0.58 &	$-$0.52 &	2.39  & 60.9$\pm$0.8& 60.8$\pm$0.5& 60.8$\pm$0.5\\
 1273&	15.796	& 0.459 & 6913 & 3.96 &	$-$0.52 &	$-$0.58 &	2.69  & 61.2$\pm$1.0& 61.5$\pm$0.6& 61.4$\pm$0.6\\
 1284&	17.284	& 0.526 & 6525 & 4.45 &	$-$0.66 &	        &	2.36  & 61.1$\pm$2.6& 61.9$\pm$2.1& 61.5$\pm$1.7\\
 1317&	17.153	& 0.508 & 6624 & 4.43 &	$-$0.57 &	        &	2.84  & 62.1$\pm$2.0& 63.4$\pm$1.5& 62.7$\pm$1.3\\
 1323&	17.134	& 0.503 & 6652 & 4.43 &	$-$0.61 &	$-$0.63 &	2.63  & 62.9$\pm$1.0& 62.7$\pm$1.2& 62.8$\pm$0.8\\
 1405&	17.263	& 0.528 & 6515 & 4.44 &	$-$0.73 &	        &	2.80  & 62.7$\pm$1.2& 61.4$\pm$3.5& 62.1$\pm$1.8\\
 1469&	17.434	& 0.545 & 6426 & 4.48 &	$-$0.42 &	$-$0.36 &	2.39  & 63.4$\pm$1.9& 62.4$\pm$1.4& 62.9$\pm$1.2\\
 1506&	16.944	& 0.499 & 6674 & 4.36 &	$-$0.50 &	        &	2.63  & 63.5$\pm$1.5& 64.3$\pm$1.8& 63.9$\pm$1.2\\
 1634&	15.734	& 0.581 & 6249 & 3.75 &	$-$0.49 &	$-$0.47 &	2.88  & 62.7$\pm$0.4& 62.0$\pm$0.5& 62.4$\pm$0.3\\
 1663&	15.729	& 0.561 & 6345 & 3.78 &	$-$0.48 &	$-$0.51 &	2.92  & 61.1$\pm$2.6& 62.1$\pm$1.8& 61.6$\pm$1.6\\ 
 1776&	17.031	& 0.512 & 6601 & 4.37 &	$-$0.74 &	$-$0.62 &	2.66  & 62.0$\pm$0.7& 62.0$\pm$0.4& 62.0$\pm$0.4\\
 1792&	17.491	& 0.537 & 6467 & 4.52 &	$-$0.66 &	        &	2.85  & 62.9$\pm$3.2& 62.9$\pm$2.0& 62.9$\pm$1.9\\
 1801&	15.632	& 0.518 & 6568 & 3.80 &	$-$0.49 &	$-$0.57 &	2.61  & 61.4$\pm$1.3& 61.1$\pm$1.6& 61.2$\pm$1.0\\
 1992&	15.141	& 0.864 & 5231 & 3.15 &	$-$0.29 &	$-$0.30 &	1.24  & 63.1$\pm$0.8& 63.1$\pm$0.5& 63.1$\pm$0.5\\

\hline				       			          
\end{tabular}
\end{center}
\end{table*}
\normalsize			       

\begin{table*}
\scriptsize
\caption{Stars with upper limits in lithium abundances.}
\label{tab:LiAbUp}
\begin{center}  
\begin{tabular}{lcccccccccc}
\hline
Object & $V$ & $V-I$ & $T_{\rm eff}$ & $\log g$ & [\ion{Fe}{i}/H] & [\ion{Fe}{ii}/H] &$\log n{\rm (Li)}$ & $V_{\rm rad}^{1}$ & $V_{\rm rad}^{2}$ & $<V_{\rm rad}>$\\
       & (mag) &  & (K) &  &  &  & & (\kms)& (\kms)& (\kms)\\
\hline

33   & 17.223 & 0.524 & 6536 & 4.43 & $-$0.58 &         & 2.05   & 61.5$\pm$1.5& 61.1$\pm$1.8& 61.3$\pm$1.2\\
80   & 16.587 & 0.475 & 6815 & 4.25 & $-$0.54 & $-$0.48 & 1.71   & 62.6$\pm$0.9& 62.7$\pm$1.2& 62.6$\pm$0.8\\
120  & 15.791 & 0.510 & 6612 & 3.88 & $-$0.50 & $-$0.59 & 2.12   & 64.9$\pm$0.7& 64.6$\pm$0.5& 64.8$\pm$0.4\\
213  & 16.371 & 0.482 & 6773 & 4.16 & $-$0.75 & $-$0.58 & 2.22   & 62.0$\pm$1.2& 62.5$\pm$1.3& 62.3$\pm$0.9\\
274  & 17.243 & 0.534 & 6483 & 4.42 & $-$0.58 & $-$0.61 & 1.79   & 63.9$\pm$1.9& 62.7$\pm$1.4& 63.3$\pm$1.2\\
354  & 16.636 & 0.478 & 6797 & 4.27 & $-$0.57 & $-$0.70 & 2.08   & 62.8$\pm$1.0& 62.0$\pm$1.2& 62.4$\pm$0.8\\
397  & 17.136 & 0.517 & 6574 & 4.41 & $-$0.66 &         & 2.13   & 63.5$\pm$2.7& 63.5$\pm$2.1& 63.5$\pm$1.7\\
439  & 16.937 & 0.487 & 6743 & 4.37 & $-$0.47 &         & 2.28   & 62.3$\pm$1.5& 63.6$\pm$1.2& 62.9$\pm$1.0\\
456  & 16.640 & 0.470 & 6845 & 4.28 & $-$0.62 &         & 1.96   & 61.9$\pm$1.0& 62.9$\pm$1.4& 62.4$\pm$0.9\\
485  & 16.870 & 0.489 & 6732 & 4.34 & $-$0.63 & $-$0.60 & 2.03   & 61.9$\pm$1.0& 61.7$\pm$1.3& 61.8$\pm$0.8\\
520  & 16.688 & 0.465 & 6875 & 4.31 & $-$0.66 &         & 2.29   & 62.7$\pm$1.9& 61.8$\pm$2.3& 62.2$\pm$1.5\\
523  & 16.274 & 0.491 & 6720 & 4.10 & $-$0.56 & $-$0.53 & 1.96   & 64.0$\pm$1.0& 64.5$\pm$0.8& 64.2$\pm$0.7\\
529  & 16.131 & 0.484 & 6761 & 4.06 & $-$0.47 & $-$0.51 & 2.09   & 62.8$\pm$1.0& 63.2$\pm$0.8& 63.0$\pm$0.6\\
547  & 16.240 & 0.486 & 6749 & 4.10 & $-$0.36 & $-$0.36 & 1.90   & 62.5$\pm$0.8& 62.5$\pm$0.6& 62.5$\pm$0.5\\
577  & 15.641 & 0.596 & 6179 & 3.69 & $-$0.39 &	$-$0.61 & 1.74   & 61.7$\pm$0.4& 61.9$\pm$0.5& 61.8$\pm$0.3\\
612  & 16.332 & 0.473 & 6827 & 4.15 & $-$0.44 & $-$0.60 & 2.26   & 62.2$\pm$1.2& 62.2$\pm$1.3& 62.2$\pm$0.9\\
631  & 15.272 & 0.770 & 5509 & 3.32 & $-$0.51 & $-$0.56 & 0.86   & 64.1$\pm$0.6& 64.2$\pm$0.4& 64.1$\pm$0.3\\
657  & 16.374 & 0.511 & 6607 & 4.11 & $-$0.68 & $-$0.67 & 2.15   & 61.0$\pm$1.0& 61.3$\pm$1.0& 61.1$\pm$0.7\\
684  & 17.033 & 0.507 & 6629 & 4.38 & $-$0.69 &         & 1.81   & 61.1$\pm$2.1& 61.6$\pm$1.7& 61.3$\pm$1.4\\
700  & 17.212 & 0.503 & 6652 & 4.46 & $-$0.52 &         & 2.03   & 61.6$\pm$2.2& 62.0$\pm$1.7& 61.8$\pm$1.4\\
770  & 16.734 & 0.480 & 6785 & 4.30 & $-$0.61 & $-$0.45 & 1.69   & 62.6$\pm$1.1& 62.4$\pm$1.2& 62.5$\pm$0.8\\
784  & 16.726 & 0.479 & 6791 & 4.30 & $-$0.56 & $-$0.50 & 1.92   & 61.1$\pm$1.3& 60.8$\pm$1.3& 61.0$\pm$0.9\\
838  & 15.869 & 0.464 & 6882 & 3.98 & $-$0.54 & $-$0.59 & 2.30   & 61.2$\pm$0.8& 61.7$\pm$0.7& 61.5$\pm$0.5\\
863  & 15.794 & 0.457 & 6925 & 3.96 & $-$0.49 &         & 2.12   & 61.9$\pm$1.1& 62.2$\pm$1.0& 62.0$\pm$0.8\\
875  & 17.190 & 0.504 & 6646 & 4.45 & $-$0.65 & $-$0.58 & 2.13   & 62.7$\pm$2.0& 62.4$\pm$1.4& 62.6$\pm$1.2\\
1035 & 16.464 & 0.458 & 6919 & 4.23 & $-$0.51 & $-$0.54 & 2.01   & 61.8$\pm$0.8& 62.1$\pm$0.8& 61.9$\pm$0.6\\
1100 & 16.390 & 0.465 & 6875 & 4.19 & $-$0.60 & $-$0.73 & 2.29   & 62.9$\pm$1.3& 62.3$\pm$1.3& 62.6$\pm$0.9\\
1253 & 16.370 & 0.486 & 6749 & 4.15 & $-$0.68 & $-$0.64 & 1.98   & 61.4$\pm$1.4& 61.8$\pm$0.9& 61.6$\pm$0.8\\
1299 & 16.868 & 0.488 & 6738 & 4.35 & $-$0.52 & $-$0.48 & 2.28   & 62.1$\pm$1.2& 62.1$\pm$1.1& 62.1$\pm$0.8\\
1321 & 16.605 & 0.470 & 6845 & 4.27 & $-$0.45 & $-$0.40 & 2.27   & 62.4$\pm$1.2& 61.8$\pm$1.0& 62.1$\pm$0.8\\
1339 & 16.454 & 0.481 & 6779 & 4.19 & $-$0.59 & $-$0.62 & 1.91   & 65.3$\pm$1.5& 65.4$\pm$1.2& 65.4$\pm$1.0\\
1375 & 16.471 & 0.476 & 6809 & 4.21 & $-$0.59 & $-$0.51 & 2.25   & 62.0$\pm$2.0& 62.4$\pm$1.4& 62.2$\pm$1.2\\
1566 & 16.375 & 0.482 & 6773 & 4.16 & $-$0.44 & $-$0.38 & 2.22   & 61.7$\pm$2.3& 61.9$\pm$1.1& 61.8$\pm$1.3\\
1590 & 16.317 & 0.495 & 6697 & 4.11 & $-$0.54 & $-$0.50 & 1.86   & 62.8$\pm$0.8& 63.0$\pm$0.8& 62.9$\pm$0.6\\
1597 & 16.567 & 0.465 & 6875 & 4.26 & $-$0.51 & $-$0.57 & 2.29   & 61.8$\pm$1.6& 62.5$\pm$1.0& 62.1$\pm$0.9\\
1635 & 17.046 & 0.499 & 6674 & 4.40 & $-$0.51 &         & 2.15   & 61.6$\pm$1.9& 60.7$\pm$1.5& 61.1$\pm$1.2\\
1648 & 15.058 & 0.855 & 5255 & 3.13 & $-$0.16 & $-$0.33 & 0.71   & 62.0$\pm$0.3& 62.1$\pm$0.2& 62.1$\pm$0.2\\
1687 & 16.520 & 0.465 & 6875 & 4.24 & $-$0.47 & $-$0.48 & 1.98   & 62.9$\pm$1.3& 62.0$\pm$0.9& 62.4$\pm$0.8\\
1709 & 16.481 & 0.463 & 6888 & 4.23 & $-$0.47 & $-$0.53 & 1.99   & 61.9$\pm$1.4& 61.8$\pm$1.1& 61.9$\pm$0.9\\
1751 & 16.461 & 0.469 & 6851 & 4.21 & $-$0.42 & $-$0.48 & 2.27   & 63.1$\pm$1.2& 62.7$\pm$0.9& 62.9$\pm$0.7\\
1752 & 16.359 & 0.493 & 6709 & 4.13 & $-$0.62 &         & 1.87   & 62.0$\pm$1.7& 61.6$\pm$1.1& 61.8$\pm$1.0\\
1839 & 16.816 & 0.481 & 6779 & 4.34 & $-$0.62 & $-$0.54 & 2.22   & 62.1$\pm$1.3& 61.0$\pm$1.5& 61.5$\pm$1.0\\
1874 & 16.496 & 0.454 & 6944 & 4.25 & $-$0.52 & $-$0.44 & 2.33   & 63.6$\pm$1.6& 62.1$\pm$1.2& 62.8$\pm$1.0\\

\hline				       			          
\end{tabular}
\end{center}
\end{table*}
\normalsize			       

\begin{table}[b]
\scriptsize
\caption{Radial velocities of the most-probable binary stars and non-members.}
\label{tab:binaries}
\begin{center}  
\begin{tabular}{lccccc}
\hline
Object & $V$ &  $V_{\rm rad}^{1}$ & $V_{\rm rad}^{2}$ & $<V_{\rm rad}>$\\
       & (mag)         &(\kms)     & (\kms)& (\kms)\\
\hline
 343 & 14.790 &    52.4$\pm$0.7&   51.8$\pm$0.7&   52.1$\pm$0.5\\
  92 & 15.751 &     15.5$\pm$0.7&   15.7$\pm$0.5&   15.6$\pm$0.4\\
 671 & 16.644 &     56.4$\pm$1.2&  109.4$\pm$2.2&   82.9$\pm$1.2\\
 182 & 15.136 &     37.2$\pm$0.4&   37.2$\pm$0.4&   37.2$\pm$0.3\\
 153 & 14.583 &     17.9$\pm$0.2&   17.7$\pm$0.2&   17.8$\pm$0.2\\
 495 & 17.268 &     96.4$\pm$2.3&   95.5$\pm$2.2&   95.9$\pm$1.6\\
1020 & 17.332 &     57.3$\pm$4.2&   55.5$\pm$5.8&   56.4$\pm$3.6\\
1906 & 14.420 &     11.4$\pm$0.7&   11.3$\pm$0.5&   11.3$\pm$0.4\\
 370 & 17.409 &     67.3$\pm$1.5&   67.7$\pm$1.8&   67.5$\pm$1.2\\
2003 & 16.872 &    64.8$\pm$5.0&   53.4$\pm$8.0&   59.1$\pm$4.7\\
2018 & 15.573 &     79.8$\pm$0.9&   80.0$\pm$0.6&   79.9$\pm$0.5\\
1438 & 16.662 &    67.6$\pm$1.5&   67.2$\pm$1.4&   67.4$\pm$1.0\\
1410 & 16.760 &      ...$^{*}$  &     ...$^{*}$ &    ...$^{*}$ \\
1631 & 16.687 &     55.8$\pm$2.8&   55.9$\pm$1.5&   55.9$\pm$1.6\\
 794 & 15.132 &     78.1$\pm$0.2&   78.0$\pm$0.6&   78.1$\pm$0.3\\
1831 & 15.566 &     44.1$\pm$0.9&   44.0$\pm$0.6&   44.1$\pm$0.5\\
1930 & 15.626 &  $-$18.0$\pm$1.1&$-$17.9$\pm$0.8&$-$17.9$\pm$0.7\\
1454 & 17.030 &    67.5$\pm$1.4&   67.5$\pm$2.0&   67.5$\pm$1.2\\
\hline		
\end{tabular}
\end{center}
$^{*}$ Almost continuum-type spectra. 		       			          
\end{table}
\normalsize			       
                                                                    
%
%

\subsubsection{Elemental abundance of other elements}

The line list used to compute the iron abundance ([Fe/H]) was taken from \cite{Pace2010}. We used this list to measure well-defined unblended \ion{Fe}{i} and \ion{Fe}{ii} lines.  As for the determination of the Li abundances,  we measured the equivalent widths of the   Fe lines (from 3 to 25 lines for \ion{Fe}{i} lines and  up to 4 lines  for \ion{Fe}{ii} lines) using the code {\it FITLINE}. We then derived elemental  abundances for the measured line equivalent widths  using {\it Turbospectrum}  \citep{Alva1998} taking the latest version of the OSMARCS atmosphere models  
(e.g., \citealt{Gusta2008}, \citealt{Plez2008}). Plane parallel 1D models were used for the computations. We adopted the 
Grevesse \& Sauval (1998) solar abundances. 

\begin{figure}
\centering
\includegraphics[angle=-90,width=0.5 \textwidth ]{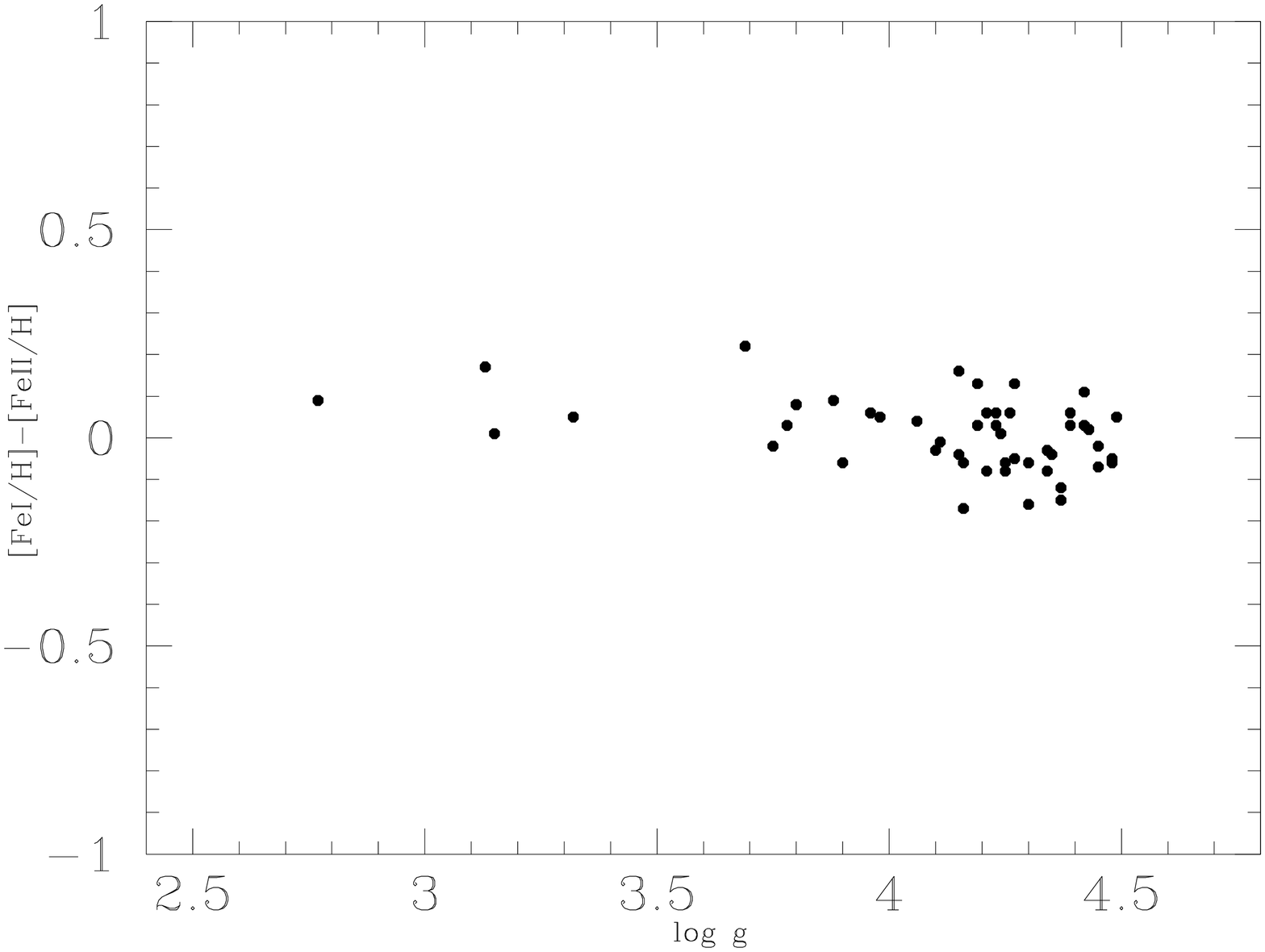}
\caption{[\ion{Fe}{i}/H] - [\ion{Fe}{ii}/H]  abundance as a function of the stellar surface gravity.} 
\label{fig:dfe_logg}
\end{figure}

 These lines have been used to compute the Fe abundance but also to check that the ionization equilibrium is consistent 
with the adopted gravity as shown in Fig.~\ref{fig:dfe_logg}. The average [Fe/H] from \ion{Fe}{i} line is of $-0.54\pm0.10$ dex, while the average from 
\ion{Fe}{ii} line is of $-0.53\pm0.14$ dex.
For the three brightest stars, we found a higher Fe abundance than  the rest of the sample. 
This result is in line with the ongoing discussion about the systematic slight differences which 
can found between Fe abundances determined in dwarfs and giants. 

We also determined the abundance of  Ca, Ti and Si  in stars where lines were detectable using the linelist of 
\cite{Pace2010}. We finally obtained  the  abundance of Ca  in 76 stars, the abundance of Ti  in 33 stars,  and the abundance of Si in 13 stars.  

For calcium, we also measured the equivalent width of four to six  lines found in this list to determine the mean [$\alpha$/Fe] for the cluster. 
The mean [Ca/Fe] abundance for NGC~2243 based on the determination of Ca in 76 stars gives [Ca/Fe]$=0.00\pm0.14$ dex, in agreement within the error bars with the slightly higher value found by \cite{grat1994}. 

 The origin of this low value is probably related to the difference in the oscillator strength determination used in
\cite{grat1994} and \cite{Pace2010}.
The aim of this work was to determine the Li abundance in a sample of stars  around the TO and it was not focused  on the determination of the [$\alpha$/Fe] ratios, so the instrumental setup was not optimized for it. 
Because of the limited wavelength range used in this study (however for a large number of stars), we could not find a single Ca line in common with other abundance studies on NGC~2243. Therefore, this solar [Ca/Fe] should be considered with caution. 
We derived a mean Ti abundance of [Ti/Fe] $=0.20\pm0.22$ dex and a Si mean abundance [Si/Fe] $=0.12\pm0.20$ dex. In conclusion, our results (i.e., $[\alpha/Fe]$ abundances ranging from 0.00 to 0.2 dex) are 
 in fair agreement with the  low $[\alpha/Fe]$ enhancement  found in other open clusters or thin disk stars of this metallicity ([Fe/H] $\simeq -0.5$ dex).

\subsection{Error estimates}

 The error in the determination of the Li abundances are a combination of the error in the measurement of the equivalent widths and the errors induced by uncertainties of the atmospheric parameters.  Errors in the determination of the equivalent widths have been estimated by changing the continuum position. The resulting errors range from 0.05 dex to 0.10 dex. 
The errors  due to the uncertainties in the atmospheric parameters have been computed by evaluating the impact on the Li abundance of a $T_{\rm eff}$ variation by 100 K, $\log g$ by 0.2 dex, and $\xi$ by 0.5 km/s on the adopted atmospheric parameters. 
These values are based on the  uncertainties found in the literature for the color excess and the distance modulus of the cluster.
An error of 0.02 on the color index corresponds to an error of $\simeq$ 100 K in the temperature scale.
An error on the distance modulus of 0.3 corresponds to a change in the surface gravity of 0.2 dex.
We added a typical uncertainty of 0.5 km/s for the microturbulent velocity.  As all the measured  lines are on the linear part of the curve of growth, the
error associated to this parameter is negligible. 

The resulting errors have been summed quadratically and give a typical error on the Li abundance of 0.14 dex. The final errors on lithium are of the order of 0.18 dex.  Errors on the Fe abundances, estimated by the line-to-line scatter, give values of 0.1 to 0.2~dex. Summing quadratically the errors
 coming from uncertainties in atmospheric parameters leads to errors ranging from 0.15 to 0.22~dex. 

The abundance uncertainties are dominated by the error in effective temperature. The chosen reddening from Kaluzny et al. (1996) is on the high side of the different 
estimations and our effective temperatures are in fact higher by about 200 K compared to the low temperature scale adopted, for instance by Hill \& Pasquini (2000). A difference of this magnitude would shift our abundances by accounting for a lower Li of about 0.15 dex and a lower [Fe/H] of 0.15 dex and a change in the ionization balance \ion{Fe}{i}  - \ion{Fe}{ii}  by 0.12 dex. Instead the effect on the [Ca/Fe] ratio is negligible. Our temperature scale is consistent with that of Kaluzny et al. (1996) ; this is not surprising 
given the adoption of the same photometry and reddening.

\section{Discussion}
\label{sec:discussion} 
\begin{figure}
\centering
\includegraphics[angle=-90,width=0.5\textwidth]{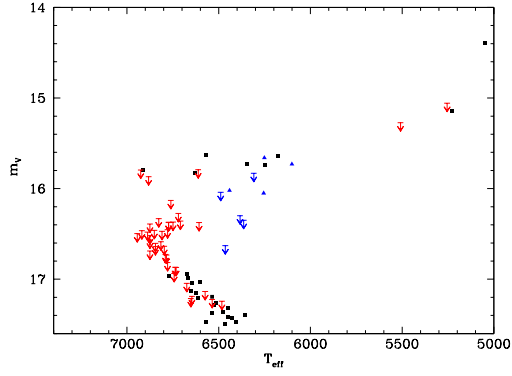}
\caption{$V$ magnitude as a function of  stellar effective temperature. Black squares are measurements for our targets. Red arrows represent upper 
limits of our Li abundance measurements. Blue symbols are results from \cite{Hill2001}, where arrows represent upper limits. } 
\label{fig:mV_Teff}
\end{figure}

\begin{figure}
\centering
\includegraphics[angle=-90,width=0.5 \textwidth ]{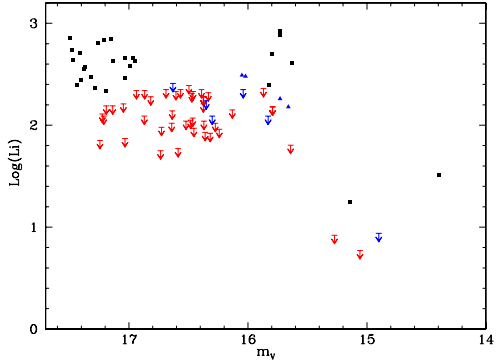}
\caption{Lithium abundance  as a function of $V$ magnitude. Symbols are the same as in Fig.~\ref{fig:mV_Teff}.}
\label{fig:Abun_Li_V}
\end{figure}

Figure~\ref{fig:mV_Teff} shows the position of the stars on a $T_{\rm eff}-V$ space, similar to a color magnitude diagram. 
It should be remembered that the \cite{Hill2001} points are on a lower 
temperature scale than that adopted. As explained above, their points would be $\sim$0.15 dex higher and $\sim$200 K moved to the 
left if they were on the same temperature scale adopted in this work. 
Fig.~\ref{fig:Abun_Li_V} shows the Li abundances as a function of the $V$ magnitude of the star. 

\subsection{The lithium dip}

The Li dip is a feature first observed  in the Hyades OC. In a narrow temperature range, the Li abundance 
is strongly depleted by factors up to 100, as first shown by  Boesgaard \& Tripicco (1986), and subsequently
 studied by several authors. \cite{Bala1995} compared several clusters in a homogeneous 
analysis concluding that the mass to which the dip occurs depends on the stellar metallicity, while 
the ZAMS $T_{\rm eff}$ does not. With NGC~2243 we have the opportunity to greatly enlarge the metallicity range explored. 

For many years the cause of the ``extra mixing'' producing the Li dip have been debated (\citealt{Talon2010} and references therein). 
Recently, Smiljanic et al. (2010) succeeded in reproducing the Li and Be 
behavior of the hot side of the Li dip in the IC~4651 cluster by using intermediate mass stars as
 low as $\sim 1.2 M_{\odot}$ and rotating models (Charbonnel \& Lagarde 2010). 

\begin{figure}
\centering
\includegraphics[angle=-90,width=0.5\textwidth]{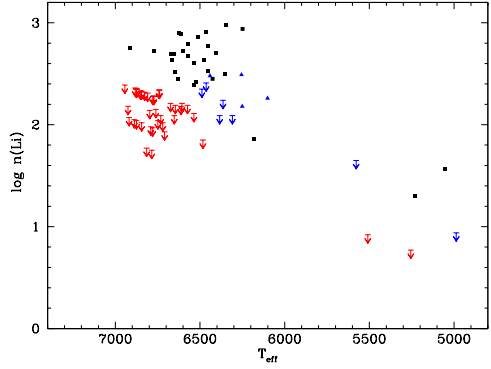}
\caption{Abundance of Li as a function of the effective temperature. Symbols are the same as in  Fig.~\ref{fig:mV_Teff}} 

\label{fig:Abun_Li_Teff}
\end{figure}

Figure~\ref{fig:Abun_Li_Teff} shows the results of the abundance determination in NGC~2243 as a 
function of the temperature of the star. This figure reveals the clear 
presence of upper limits in a narrow range of temperature starting around $T_{\rm eff}$ = 6700 K. 
The range of temperature $6700-6900$ K represents the Li dip 
position for this cluster. When informations from Fig.~\ref{fig:Abun_Li_Teff} and Fig.~\ref{fig:mV_Teff}
 are put together, one can see that the stars without a Li 
detection are located in the upper (hottest) part of the ZAMS. Some non detections are also found for 
cooler (so fainter) stars. We attribute the lack of detection in the faint stars 
to the resulting low $S/N$ ratio in their spectra. Star 1273, with a temperature $T_{\rm eff}$= 6913 K and a magnitude of 
$V=15.8$ mag, shows a Li abundance of 2.70 dex and seems to be the first star located just on the blue side of the dip. 

Once proven beyond any doubt the presence of the dip in this metal-poor cluster, we may discuss some
 additional points. The effective temperature of the dip is higher than that observed in other clusters. 
 The mass of the dip can be estimated with the help of evolutionary isochrones. Using the Girardi 
(web page\footnote{http://stev.oapd.inaf.it/~lgirardi/}) tracks, a cluster with a turnoff of 6900 K and a 
metal content of [Fe/H]=$-0.7$ is expected to have an age of 3.5 Gyr and a turnoff mass of $1.12 M_\odot$. 
Using the same isochrone, we could derive approximative masses for the eclipsing binary studied by 
Kaluzny et al. (2006), finding 0.97 $M_\odot$ for their components, in fair agreement with the 
estimated mass by Kaluzny et al. (2006) of  1.089 and 1.069 $M_\odot$. Positioning the red side
 of the Li dip at 6700 K, the expected mass of the red edge of the dip would be of 1.03 $M_\odot$. 

The Li dip in this metal-poor cluster is therefore characterized by higher temperatures and substantially
 lower masses compared to  more metal-rich clusters. 
In the Hyades and Praesepe clusters, the dip center is at about 1.4 $M_\odot$ (Balachandran 1995). In IC~4651, with [Fe/H]=0.11, Smiljanic et al. (2010) did not 
compute the mass of the center, but estimated the red edge of the dip to be at 1.2 $M_\odot$. In general, the decrease of the Li dip mass with metallicity is 
very clear, and the effect is clearly visible in NGC~2243, given its low metallicity. 
Adopting a temperature scale cooler by $\simeq 200 K$ degrees would move the dip towards cooler stars, which means also lower masses. Thus the use of a cooler  
$T_{\rm eff}$ scale would not change this result.

%
%

\begin{figure}
\centering
\includegraphics[angle=-90,width=0.5\textwidth]{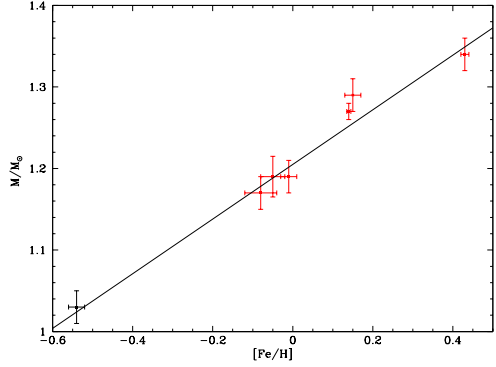}
\caption{ Correlation of the dip cool-side position in mass and the cluster's [Fe/H] ratio.  The black symbol
represents our result for NGC2243; red symbols are data from the literature. The black line represents a linear
regression fit to the data. The slope value is 0.33 and the correlation coefficient is 0.98.} 
\label{fig:mdip_feh}
\end{figure}

Figure~\ref{fig:mdip_feh} shows the position of the dip cool-side  expressed in mass  as a function of the metallicity [Fe/H] for a sample of OCs (\citealt{Cum2012} and references therein). 
With the addition of NGC~2243, the [Fe/H] range covered has a metallicity of almost 1 dex.
This figure reveals a rather tight correlation over the whole range of metallicity.
This dependence of the Li dip mass from the metallicity seems to be a well-established feature that models should reproduce.

\subsection{Li chemical evolution} 
Observations of many clusters have shown that the regions in the color diagram at the edge of the Li dip show the highest level of Li abundances, 
and probably indicate the Li abundance of the pristine gas. 
NGC~2243 is not an exception and the five stars bluer to the dip show an average Li abundance of 2.70 $\pm0.20$ dex. 
We may assume this value to be a conservative lower limit of the pristine Li abundance in the cluster. This value is consistent, within errors, with 
the upper envelope of Li abundances observed in thin disk stars of metallicity similar to that of NGC~2243. 
\cite{lambert2004} in their Table 2 provide
the mean Li abundance for the six stars richest  in Li in each metallicity bin and for the bin --0.4 to --0.6 this is 2.64$\pm 0.07$ dex. The recent large sample of 
\cite{rami2012} is consistent with this value. The models for Li evolution of \cite{Prantzos2012}  predict a value of Li at this metallicity that is 
about a factor of two higher. Thus, in order to make these models compatible with the observations, it is necessary to assume that even the stars with the 
highest observed Li abundances at this metallicity have depleted Li by a factor of two. This is a direct consequence of the fact that these models assume 
that the primordial level of Li is provided by the baryonic density measured from the fluctuations of the Cosmic Microwave Background by the WMAP satellite 
\citep{LDH11,KSD11}, coupled with standard Big Bang nucleosynthesis. The same models, assuming the level of the Spite plateau 
\citep{spite1982,SboBC10} as primordial Li abundance, would be in good agreement with the upper envelope of Li abundances in 
thin disk stars at all metallicities, without the need of extra Li depletion. For a review on the evolution of Li see also \cite{mat10}.
\cite{Howk2012} estimate from the insterstellar medium (ISM)  the Li abundance of the Small Magellanic Cloud equal to   $2.68 \pm 0.16 $ dex for a metallicity of [Fe/H]=$-0.59$ dex.
Our  estimate is  $\log n{\rm (Li)}= 2.70 \pm 0.20$ dex for [Fe/H] = $-0.54$ dex, therefore in close agreement with the ISM measurements  at the same metallicity. 
Our results would tend to confirm the finding by Howk et al. (2012) that the Li evolution is extremely constrained when the Li primordial abundance from WMAP
is assumed.

\subsection{Comparison with 47 Tuc}

As already mentionned in the introduction, one of the interesting characteristics of NGC~2243 is that its metallicity almost overlaps with that of the well-studied globular cluster 47 Tuc. 

The highest Li abundance in NGC 2243 is, as in all young OCs, found at the blue edge of the Li dip and is 
of  2.70 $\pm$ 0.2 dex, with values potentially as high as 2.9 dex.  47 Tuc has been studied in the past by several groups.
It was first observed  by \cite{pas97} who found three stars at a level of $\log n{\rm (Li)}$=2.35 dex, but these observations, 
made with the 3.5m NTT telescope, had a low $S/N$ ratio. With the advent of the VLT, higher quality observations and more objects became available. 
\cite{bon07} observed four stars, finding evidence of strong  Li variability and a possible correlation with other light elemental abundances, 
typical of globular clusters. This result was later confirmed by \cite{doraz10} who have used FLAMES/GIRAFFE data for  one hundred stars, showing that a large Li variation is present. 
Interestingly, this Li variation is not related to the Na-O correlation found in this cluster.

The extensive work of D'Orazi et al. (2010) confirms the presence of a very large Li variation in this cluster, with  Li between 2.4 and 2.5 dex and one object as high as 2.78 dex. Moreover, the maximum observed Li abundance 
in  47~Tuc stars is about 0.2 dex lower than in NGC 2243.  It is not easy, however, to firmly establish whether these abundances are representative of the pristine Li abundance of the cluster gas. While this is the case for NGC2243 (stars bluer than the gap do not show evidenc, either observational or theoretical, for Li depletion),  this is not the case for 47 Tuc. The turnoff stars of this cluster are rather cool (T$_{\rm{eff}}$ less than $\simeq$ 5800 K), more than 500 degrees K cooler than the faintest stars observed in NGC2243.  At this effective temperature, old OC stars already show evidence of main sequence Li depletion
\citep{pasq00,Sestito05}. It is important to note that none of the OCs studied so far  is as metal-poor as 47 Tuc or NGC~2243; it would be very important to observe fainter stars in NGC 2243 to confirm the Li trend  with effective temperature. Still, the possibility that the Li observed in 47 Tuc stars does not correspond to the pristine Li, but to slightly depleted Li, cannot be excluded. 
Keeping this possibility in mind, it is worth analyzing the case that the Li abundance in the two clusters was not initially the same.

Given its old age, in spite of its  high 
metallicity, it would be natural to assume that 47 Tuc started from the primordial Li abundance, $\log n{\rm (^{7}Li)} = 2.72$ dex, from WMAP+SBBN 
\citep{CFO08}. Instead, NGC~2243, with a much younger age, should have incorporated the Li produced by the Galactic sources, 
thus $\log n{\rm (^{7}Li)} \sim 3.0$ dex, according to the models of \cite{Prantzos2012} at this metallicity. If this is the case, one has to conclude 
that the Li depletion with time is a highly non-linear process. In fact, the two clusters started with a difference in Li of a factor of 2 
and they retain this difference, in spite of the fact that in 47 Tuc the Li has had 8 Gyrs more to accomplish its depletion. Another intriguing aspect 
of this scenario is that the amount of depletion (initial/present) is again roughly a factor of two in both cases. The conclusion could be that Li 
is depleted very rapidly, by a factor of two, from whatever value it starts from, and then remains constant for all of the star's lifetime
without taking into account the ``extra mixing'' that enters into play to explain the ``Li dip'' and the cooler stars. This scenario could be 
consistent with the suggestion that the Li depletion takes place during the pre-main sequence  (\cite{molaro2012}.
However, the observational evidences show that pre-main Sequence (PMS) depletion in solar stars is not really effective (Randich et al. 2001; 
Jeffries 2006). Considering the relatively low metal abundance of these clusters compared to the Sun, even less PMS depletion would be theoretically expected.

This promising picture is, however, troubled if we note that the star with the highest level of Li abundance in the \cite{doraz10} sample has 
$\log n{\rm (^{7}Li)}=2.78$, i.e.,  close to the highest value also observed in NGC~2243.  This ``upper limit" of the Li content in 47 Tuc stars  can be considered  the closest value  to the original Li content of the cluster. 

This value would then suggest that the two clusters started with similar Li abundance. The implications of this are puzzling. We see three possible situations:
\begin{enumerate}
\item Li is depleted both in 47 Tuc and NGC~2243, 47 Tuc already benefitted from an initial Li that was a factor of two higher than the 
primordial value and thereafter the Li abundance was constant up to the time of the NGC~2243 cluster formation; 
\item neither 47 Tuc nor NGC~2243 have depleted any Li and were formed with the primordial abundance. Again no Li is produced 
between 12 Gyr ago and 4 Gyr ago;
\item 47 Tuc started with the primordial Li abundance, but its highest Li stars suffered no Li depletion. NGC 2243 started with a 
higher Li-abundance and its highest Li stars suffered depletion by a factor of two.  
\end{enumerate}

All of these contradict our current understanding of Li evolution and stellar structure.  It would certainly be useful to 
extend the work of \cite{doraz10}  and have new determinations of the  Li abundance  in 47 Tuc, to check if  this very high  Li abundance 
found in one star is  real and is not simply a spurious result due to a low S/N spectrum or incorrect stellar parameters. 

If the Li initial abundance were the same between the two clusters, then the Li enrichment would  basically be determined by the Fe content rather than by the star formation history of the two clusters, which was quite different, as shown by their different [$\alpha$/Fe] enrichment. 

\section{Conclusion}

\label{sec:concl}

{VLT-FLAMES observations of a sample of stars belonging to NGC2243 has permitted us to determine the Li 
abundances in 27 stars   and the Fe and Ca abundances in 76 stars.  We found  a metallicity of [Fe/H] =
$-0.54 \pm 0.10$ dex and  [Ca/Fe] = $0.00 \pm 0.14$ dex.  The Li dip is well defined with a center 
corresponding to a mass of $M = 1.06~M_{\odot}$, a value much smaller than the value found for solar 
metallicity OCs. 
We confirm a correlation between the Li dip cool-side position in mass as a function of the cluster's [Fe/H] ratio, 
extending the relation towards metallicity down to [Fe/H] $\simeq -0.55 $ dex.
A Li abundance of $2.70 \pm 0.2$ dex found at the blue edge of the Li dip of  NGC 2243 is in good agreement with results obtained  for the Li of  the interstellar medium in the SMC  having similar low metallicity. This value of Li in metal-poor objects is comparable to the Li abundance deduced from WMAP measurements putting strong constraints on the models of Li enrichment during the early history of our Galaxy.}

\begin{acknowledgements}

The authors are grateful to the anonymous referee for a very careful reading of the paper and valuable comments. 
P.F. acknowledges support from the ESO Visitor program. 
K.B. was supported by the ESO DGDF 2008, and by the Italian {\em Ministero dell'Istruzione, Universit\`a e Ricerca} (MIUR) fundings. 
We thank Simone Zaggia for the discussions about the GIRAFFE code girBLDRS. PF and PB acknowledge support from the Programme National 
de Physique Stellaire (PNPS) and the Programme National de Cosmologie et Galaxies (PNCG) of the Institut National de Sciences de 
l'Universe of CNRS. This research made use of WEBDA, operated at the Institute for Astronomy of the University of Vienna, and of 
SIMBAD/VIZIER databases operated at the CDS (Strasbourg, France).

\end{acknowledgements}

\bibliographystyle{aa}

\end{document}